\documentclass[twocolumn,showpacs,aps,prd,]{revtex4-1}

\usepackage{graphicx}
\usepackage{dcolumn}
\usepackage{amsmath}
\usepackage{epsfig}
\usepackage{subfigure}
\usepackage{epstopdf}
\usepackage{multirow}
\usepackage{mathrsfs}
\usepackage[colorlinks,urlcolor=blue,citecolor=blue,linkcolor=blue] {hyperref}
\usepackage{lineno}
\usepackage{enumerate}
\usepackage{float}

\begin{document}


\title{\boldmath Search for the lepton number violating decay
  $\Sigma^{-} \to p e^{-} e^{-}$ and the rare inclusive decay $\Sigma^{-} \to \Sigma^{+} X$ }

\author{
\begin{small}
\begin{center}
M.~Ablikim$^{1}$, M.~N.~Achasov$^{10,c}$, P.~Adlarson$^{67}$, S. ~Ahmed$^{15}$, M.~Albrecht$^{4}$, R.~Aliberti$^{28}$, A.~Amoroso$^{66a,66c}$, M.~R.~An$^{32}$, Q.~An$^{63,49}$, X.~H.~Bai$^{57}$
% ...
\\
\vspace{0.2cm}
(BESIII Collaboration)\\
\vspace{0.2cm} {\it
$^{1}$ Institute of High Energy Physics, Beijing 100049, People's Republic of China\\
% ...
$^{4}$ Bochum Ruhr-University, D-44780 Bochum, Germany\\
% ...
$^{10}$ G.I. Budker Institute of Nuclear Physics SB RAS (BINP), Novosibirsk 630090, Russia\\
% ...
$^{15}$ Helmholtz Institute Mainz, Staudinger Weg 18, D-55099 Mainz, Germany\\
% ...
$^{28}$ Johannes Gutenberg University of Mainz, Johann-Joachim-Becher-Weg 45, D-55099 Mainz, Germany\\
% ...
$^{32}$ Liaoning Normal University, Dalian 116029, People's Republic of China\\
% ...
$^{49}$ State Key Laboratory of Particle Detection and Electronics, Beijing 100049, Hefei 230026, People's Republic of China\\
% ...
$^{57}$ University of Jinan, Jinan 250022, People's Republic of China\\
% ...
$^{63}$ University of Science and Technology of China, Hefei 230026, People's Republic of China\\
% ...
$^{66a}$ University of Turin and INFN, University of Turin, I-10125, Turin, Italy\\ 
$^{66b}$ University of Turin and INFN, University of Eastern Piedmont, I-15121, Alessandria, Italy \\ 
$^{66c}$ University of Turin and INFN, INFN, I-10125, Turin, Italy\\
$^{67}$ Uppsala University, Box 516, SE-75120 Uppsala, Sweden\\
% ...
$^{c}$ Also at the Novosibirsk State University, Novosibirsk, 630090, Russia\\
% ...
}
\end{center}
\vspace{0.4cm}
\end{small}
}


\begin{abstract}
Using a data sample of $(1310.6 \pm 7.0) \times 10^{6}$ $J/\psi$
events taken with the BESIII detector at the center-of-mass energy of
3.097 GeV, we search for the first time for the lepton number
violating decay $\Sigma^{-} \to p e^{-} e^{-}$ and the rare inclusive
decay $\Sigma^{-} \to \Sigma^{+} X$, where $X$ denotes any possible
particle combination. The $\Sigma^-$ candidates are tagged in $J/\psi
\to \bar{\Sigma}(1385)^+\Sigma^-$ decays.  No signal candidates are
found, and the upper limits on
the branching fractions at the 90\% confidence level are determined to
be $\mathcal{B}(\Sigma^{-} \to p e^{-} e^{-}) < 6.7 \times 10^{-5}$
and $\mathcal{B}(\Sigma^{-} \to \Sigma^{+} X) < 1.2 \times 10^{-4}$.
\end{abstract}


\maketitle

\section{\boldmath INTRODUCTION}
In the Standard Model (SM)~\cite{Glashow:1961tr, Salam:1964ry,
  Weinberg:1967tq} of particle physics, lepton number conservation is
associated with a global $U(1)_{L}$ symmetry. In addition, under the
postulate of massless neutrinos, $U(1)_{e} \times U(1)_{\mu} \times
U(1)_{\tau}$ is an automatic global symmetry, which means that
individual lepton-flavor numbers --- $e$-number, $\mu$-number, and
$\tau$-number --- are expected to be conserved. However, the
discoveries of neutrino oscillations~\cite{Fukuda:1998fd,
  Fukuda:1998mi, Ahmad:2002jz, Eguchi:2002dm}, the matter anti-matter
asymmetry of the universe~\cite{Trodden:1998ym, Dine:2003ax,
  DiBari:2013rga, Sakharov:1967dj} and the existence of dark
matter~\cite{White:1992ri, Allen:2011zs, Salucci:2018hqu} require new
physics theories beyond the SM. New physics models of non-zero
neutrino masses predict neutrinos to be Dirac or Majorana
fermions~\cite{Mohapatra:1979ia, Schechter:1980gr, Cheng:1980qt,
  Ma:1998dn}. If neutrinos are Dirac fermions, $U(1)_{L}$ may remain
as an exact global symmetry. However, if neutrinos are Majorana
fermions, $U(1)_{L}$ is not a good global symmetry. Currently, we
cannot distinguish whether neutrinos are Dirac fermions or Majorana
fermions. Hence, it is important to investigate the validity of
lepton-number conservation directly.  Observation of lepton number
violating (LNV) processes would explicitly point out the direction of
new physics, while experimental upper limits (ULs) could translate
into stringent conditions for theoretical models.

\begin{figure}[!htbp]
\centering
\includegraphics[width=0.38\textwidth]{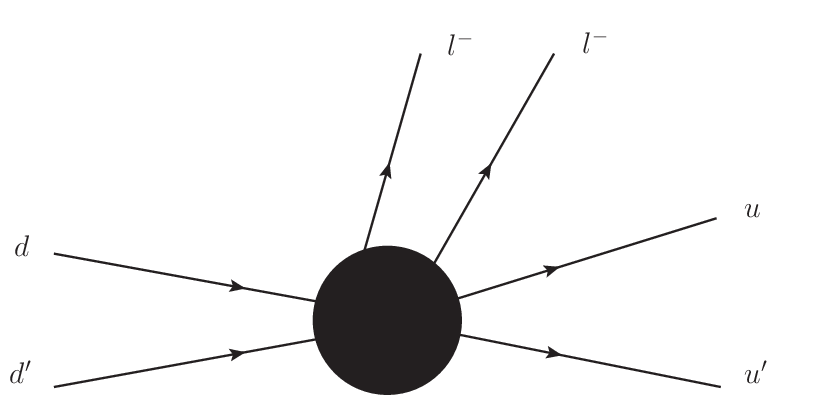}
\caption{Typical Feynman diagram for $B_{1}^{-} \to
  B_{2}^{+}l^{-}l^{-}~(l=e,\mu)$, in which two down type quarks
  convert into two up type quarks and two leptons.}
\label{diagram1}
\end{figure}

\begin{figure}[!htbp]
\centering
\includegraphics[width=0.38\textwidth]{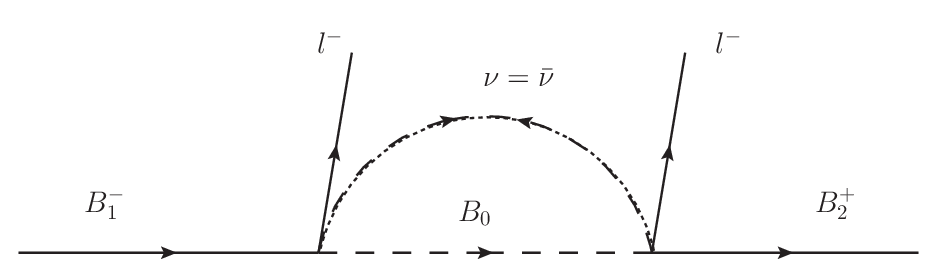}
\caption{Feynman diagram for $B_{1}^{-} \to
  B_{2}^{+}l^{-}l^{-}~(l=e,\mu)$, where a loop of a virtual baryon,
  $B_{0}$, and a Majorana neutrino, $\nu$, is introduced.}
\label{diagram2}
\end{figure}

A number of experiments have searched for LNV in meson
decays~\cite{Zyla:2020zbs}, while only a few experiments have
reported searches in hyperon decays~\cite{Littenberg:1991rd,
  Rajaram:2005bs}. The LNV decay of $B_{1}^{-} \to
B_{2}^{+}l^{-}l^{-}~(B = {\rm{baryon;~}} l =e,\mu)$ is a unique
process, in which two down-type ($d$ or $s$) quarks convert into two
up-quarks changing the charge of the hyperons according to the $\Delta
Q = \Delta L = 2$ rule, where $\Delta Q$ and $\Delta L$ are the
changes of charge number and lepton number, respectively. The
transition of the quarks is assumed to occur at the same space-time
location, as shown in Fig.~\ref{diagram1}, and is determined by local
four-quark operators~\cite{Barbero:2002wm, Barbero:2007zm,
  Barbero:2013fc}. The underlying mechanism is similar to that of
neutrinoless double beta $(0\nu\beta\beta)$ nuclear decay $(A, Z) \to
(A, Z+2) e^{-}e^{-}$, which is a sensitive probe in the search for the
effects of very light Majorana neutrinos~\cite{Rodejohann:2011mu,
  Dolinski:2019nrj}. In Refs.~\cite{Barbero:2002wm, Barbero:2007zm},
based on a model where the dominant contributions are given by a loop
of a virtual baryon and a Majorana neutrino, as shown in
Fig.~\ref{diagram2}, the predicted branching fractions of $\Sigma^{-}
\to p e^{-} e^{-}$ and $\Sigma^{-} \to \Sigma^{+} e^{-} e^{-}$, can
reach $10^{-31}$ and $10^{-35}$, respectively. While in
Ref.~\cite{Barbero:2013fc}, based on the Massachusetts Institute of
Technology (MIT) bag model~\cite{Chodos:1974je, Chodos:1974pn}, the
branching fractions are increased by several orders of magnitude, and,
for example, the branching fraction of $\Sigma^{-} \to p e^{-} e^{-}$
can reach $10^{-23}$.

In this paper, using the process $J/\psi \to \bar{\Sigma}(1385)^+
\Sigma^{-}$ ~\cite{Zyla:2020zbs} from the data sample of $(1310.6
\pm 7.0) \times 10^6$ $J/\psi$ events~\cite{Ablikim:2012cn,
  Ablikim:2016fal, Li:2016tlt} collected with the BESIII detector, we
present the first search for the $\Delta Q = \Delta L = 2$ process
in $\Sigma^{-}$ decays. In the channel $\Sigma^{-} \to \Sigma^{+}
e^{-} e^{-}$, due to the limited phase space ($M_{\Sigma^{-}} -
M_{\Sigma^{+}} \simeq 8 ~{\rm{MeV/}}c^{2}$) and the small momentum of the
$\Sigma^{-}$, the leptons have very small momenta and cannot be
reconstructed in the detector. Therefore, the processes investigated
in this analysis are $\Sigma^{-} \to p e^{-} e^{-}$ and the rare
inclusive decay $\Sigma^{-} \to \Sigma^{+} X$, where $X$ represents
any particles or particle combinations, including $ e^{-}
e^{-}$. Throughout this paper, the charge conjugate channels
$\bar{\Sigma}^{+} \to \bar{p} e^{+} e^{+}$ and $\bar{\Sigma}^{+} \to
\bar{\Sigma}^{-} \bar{X}$ are investigated at the same time.
It is important to note that the blind analysis method is used in this paper.
The Monte Carlo sample is used to determine the analysis strategy.
Then, with fixed strategy, the data sample is opened to obtain the final results.

\section{\boldmath BESIII DETECTOR AND MONTE CARLO SIMULATION}\label{sec:secII}
The BESIII detector is a magnetic spectrometer~\cite{Ablikim:2009aa} located at the Beijing Electron Positron Collider (BEPCII)~\cite{Yu:2016cof}. The cylindrical core of the BESIII detector consists of a helium-based multilayer drift chamber (MDC), a plastic scintillator time-of-flight system (TOF), and a CsI(Tl) electromagnetic calorimeter (EMC), which are all enclosed in a superconducting solenoidal magnet providing a 1.0~T (0.9~T in 2012) magnetic field. The solenoid is supported by an octagonal flux-return yoke with resistive plate counter muon identifier modules interleaved with steel. The acceptance of charged particles and photons is 93\% over $4\pi$ solid angle. The charged-particle momentum resolution at $1~{\rm{GeV}}/c$ is $0.5\%$, and the $dE/dx$ resolution is $6\%$ for the electrons from Bhabha scattering. The EMC measures photon energies with a resolution of $2.5\%$ ($5\%$) at $1$~GeV in the barrel (end cap) region. The time resolution of the TOF barrel part is 68~ps, while that of the end cap part is 110~ps.

Simulated samples produced with the {\sc
  geant4}-based~\cite{Agostinelli:2002hh} Monte Carlo (MC) software,
which includes the geometric description of the BESIII detector and
the detector response, are used to determine the detection efficiency
and to estimate the backgrounds. The simulation includes the beam
energy spread and initial state radiation (ISR) in the $e^+e^-$
annihilations modeled with the generator {\sc
  kkmc}~\cite{Jadach:1999vf,Jadach:2000ir}.

The inclusive MC sample consists of the production of the $J/\psi$
resonance, and the continuum processes ($e^{+} e^{-} \to q \bar{q}$)
incorporated in {\sc kkmc}~\cite{Jadach:1999vf,Jadach:2000ir}. The
known decay modes of $J/\psi$ are modeled with {\sc
  evtgen}~\cite{Lange:2001uf, Ping:2008zz} using branching fractions
taken from the Particle Data Group~\cite{Zyla:2020zbs}, and the
remaining unknown decays from the charmonium states with {\sc
  lundcharm}~\cite{Chen:2000tv, Yang:2014vra}. The final state
radiation (FSR) from charged final state particles are incorporated
with the {\sc photos}~\cite{RichterWas:1992qb} package.

\section{\boldmath EVENT SELECTION}
In this paper, the $\Sigma^-$ data sample is obtained through the
process $J/\psi \to \bar \Sigma(1385)^{+} \Sigma^{-}$. A double-tag
method, which was developed by the MARK-III
experiment~\cite{Baltrusaitis:1985iw}, is employed to determine the
absolute branching fraction and reduce the systematic uncertainties.
First, we reconstruct $\bar \Sigma(1385)^{+}$ via the decay $\bar
\Sigma(1385)^{+} \to \bar \Lambda \pi^+$ and then determine the
number of $\Sigma^{-}$ events in the recoil mass spectrum of the
$\bar \Sigma(1385)^{+}$, which is defined in Eq.(\ref{equ:recoilM}). 
These events are referred to as `single tag'
(ST) events.  Next, we search for signal candidates in the selected
$\Sigma^{-}$ sample by looking directly for their decay products. Events
with signal candidates are called `double tag' (DT) events. The absolute
branching fraction is calculated by
\begin{equation}
\mathcal{B}_{\rm{sig}}=\frac{N_{\rm{DT}}^{\rm{obs}}}{N_{\rm{ST}}^{\rm{obs}}\epsilon_{\rm{DT}} /\epsilon_{\rm{ST}}},
\label{equ:BF}
\end{equation}
where $N_{\rm{ST}}^{\rm{obs}}$ is the ST yield,
$N_{\rm{DT}}^{\rm{obs}}$ is the DT yield, $\epsilon_{\rm{ST}}$ and
$\epsilon_{\rm{DT}}$ are the ST and the DT efficiencies.

\subsection{\boldmath ST event selection}
In the selection of ST events, $\bar \Sigma(1385)^{+}$ is
reconstructed via the $\bar \Sigma(1385)^{+} \to \bar \Lambda \pi^+$
decay. All charged tracks are required to have a polar angle within
$|\cos\theta|<0.93$. The $\bar \Lambda$ is reconstructed via $\bar
\Lambda \to \bar p \pi^{+}$ decay. Each track used to reconstruct
$\bar \Lambda$ is required to have a distance of closest approach to
the interaction point (IP) along the beam direction less than
20~cm, while the bachelor pion candidates are required to have a
distance of closest approach to the IP less than 1~cm in the plane
perpendicular to the beam and less than 10~cm along the beam
direction. The values of 20~cm, 1~cm and 10~cm are based on track performance study.

We perform particle identification (PID) on the charged tracks with
the information of $dE/dx$ measured in the MDC and the time of flight
measured by the TOF. The confidence levels (CLs) for the pion, kaon,
and proton hypotheses ($CL_{\pi}$, $CL_{K}$, and $CL_{p}$) are
calculated. The anti-proton candidates are required to satisfy $CL_{p}
> 0.001$, $CL_{p} > CL_{\pi}$, and $CL_{p} > CL_{K}$. The bachelor
pion candidates are required to satisfy $CL_{\pi} > 0.001$ and
$CL_{\pi} > CL_{K}$, while there is no PID requirement for the pion
from $\bar \Lambda$ decay.

The two charged tracks used to reconstruct $\bar \Lambda$ are
constrained to originate from a common decay vertex by performing a
primary vertex fit on the two tracks. The $\chi_{1}^{2}$, which
represents the goodness of the primary vertex fit, is required to be
less than 100. A secondary vertex fit is also performed on the
same daughter tracks of $\bar \Lambda$ candidates, imposing the additional
constraint that the momentum of the candidate points back to the
IP. The $\chi_{2}^{2}$ of the secondary vertex fit is required to be
less than 100. The two values of $\chi^{2}$ requirements result in  high quality vertex fits.
To further suppress non-$\bar{\Lambda}$ background,
the decay length of $\bar \Lambda$, which is the distance between the
IP and the secondary vertex, is required
to be larger than 2 standard deviations of the decay length. The
fitted four-momentum of the $\bar p \pi^{+}$ combination is used in
further analysis, and the invariant mass of the $\bar p \pi^{+}$
combination is required to be within $(1.112,
1.120)$~${\rm{GeV}}/c^{2}$.

The recoil mass of $\bar \Sigma(1385)^{+}$ is defined as 
\begin{equation}
\footnotesize
  M_{\rm{recoil}} =  \sqrt{ \left( E_{\rm{J/\psi}} - E_{\bar \Lambda} - E_{\pi^{+}} \right)^{2} - \left( \vec{\mathbf{p}}_{\rm{J/\psi}} - \vec{\mathbf{p}}_{\bar \Lambda} - \vec{\mathbf{p}}_{\pi^{+}} \right)^{2} },
\label{equ:recoilM}
\end{equation}
where, $E_{\rm{J/\psi}}$$\left( \vec{\mathbf{p}}_{\rm{J/\psi}} \right
)$, $E_{\bar \Lambda}$$\left( \vec{\mathbf{p}}_{\bar \Lambda} \right)$
and $E_{\pi^{+}}$$\left( \vec{\mathbf{p}}_{\pi^{+}} \right)$ are the
energies (momenta) of $J/\psi$, $\bar \Lambda$ and $\pi^{+}$ in the
$J/\psi$'s center-of-mass frame. To suppress backgrounds, such as
$J/\psi \to \bar \Sigma(1385)^{+} \Sigma(1385)^{-}$, the sum of
$E_{\bar \Lambda}$ and $E_{\pi^{+}}$ is required to be within $(1.59,
1.70)$~$\rm{GeV}$.

\begin{figure}[!htbp]
\centering
\includegraphics[width=0.38\textwidth]{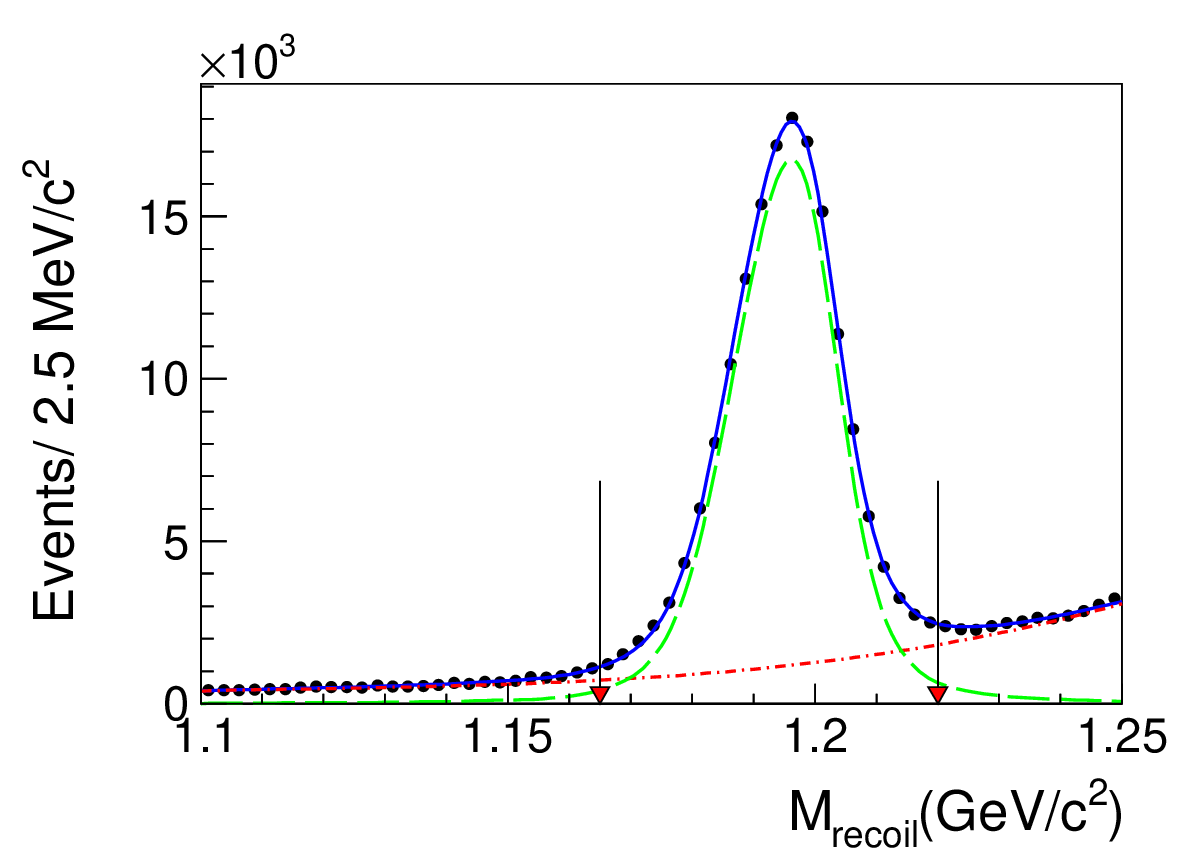}
\caption{Fit to the $M_{\rm{recoil}}$ distribution of ST events in
  data. Points with error bars represent data. The solid blue line is
  the total fit, the green line is the signal shape and the red line is
  the background component. The arrows denote the $M_{\rm{recoil}}$
  signal region.}
\label{fig:recoilingmass}
\end{figure} 

All $\bar \Sigma(1385)^{+}$ candidates in
an event are retained. We then fit the
$M_{\rm{recoil}}$ distribution to obtain the ST
yield. Figure~\ref{fig:recoilingmass} shows the fit to the
$M_{\rm{recoil}}$ distribution of data. In the fit, the
background is described by a second order Chebychev polynomial
function, and the signal shape is modeled by MC simulated shape
convolved with a Gaussian function to account for the resolution
difference between data and MC simulation. The mean and
the width of the Gaussian function are additional free parameters in
the fit. The ST
efficiency obtained from MC simulation is $(31.59\pm0.09)\%$. With the ST yield
returned by the fit, $N_\text{ST} = 147743\pm563$, we obtain $\mathcal{B}(J/\psi
\to \bar \Sigma(1385)^{+}\Sigma^-)=(3.21\pm0.07)\times10^{-4}$, where
the uncertainty is statistical only. This branching fraction is
compatible with the world average value taken from the PDG~\cite{Zyla:2020zbs}, 
$(3.1\pm0.5)\times10^{-4}$, within the large uncertainties of the
world average.

\subsection{\boldmath DT event selection}
In the recoil side of the selected ST events, we search for the LNV
process $\Sigma^- \to p e^- e^-$ and the rare inclusive decay
$\Sigma^- \to \Sigma^+ X$, using the charged tracks and
electromagnetic showers not used previously. Each
charged track is also required to have a polar angle within
$|\cos\theta|<0.93$ and a distance of closest approach to the
interaction IP along the beam direction less than 20~cm. The momentum
of the $\Sigma^{-}$ is small and the phase space in $\Sigma^{-} \to
\Sigma^{+} X$ is extremely small like that in $\Sigma^{-} \to
\Sigma^{+} e^{-} e^{-}$ due to the small difference in the $\Sigma^{\pm}$
masses.  Therefore the momenta of particles in the $\Sigma^{-} \to
\Sigma^{+} X$ decay, except for $\Sigma^{+}$ reconstructed via
$\Sigma^{+} \to p \pi^{0}$, are too small to reach the MDC and other
detectors. So only three charged tracks are required for $\Sigma^- \to
p e^- e^-$ and one charged track for $\Sigma^- \to \Sigma^+ X$.

Proton PID is performed as above. Electron PID is performed
using the $dE/dx$, TOF, and EMC information, with which the CLs for
electron, pion and kaon hypotheses ($CL_{e}$, $CL_{\pi}$ and 
$CL_{K}$) are calculated. Electron candidates are
required to satisfy $C L_{e} > 0.001$ and $C L_{e}/(C L_{e}+C
L_{\pi}+C L_{K})>0.8$.

Electromagnetic showers are reconstructed from clusters of energy
deposited in the EMC. The photon candidate showers must have a minimum
energy of 25~MeV in the barrel region $(|\cos\theta|<0.80)$ or 50~MeV
in the end cap regions $(0.86<|\cos\theta|<0.92)$. To suppress
electronic noise and energy deposits unrelated to the event, timing
information from the EMC for the photon candidates must be in
coincidence with collision events, with a requirement of $0 \leq t \leq
700$~$\rm{ns}$. The $\pi^{0}$ candidates are reconstructed from pairs
of photon candidates. Due to the worse resolution in the end cap regions of the
EMC, $\pi^{0}$ candidates reconstructed with both photons in the end
caps of the EMC are rejected. The invariant mass of two photons is
required to be within $(0.115, 0.150)$~${\rm{GeV}}/c^{2}$ for
$\pi^{0}$ candidates. To improve the overall kinematic resolution, a
mass-constraint kinematic fit is performed by constraining the $\gamma
\gamma$ invariant mass to the nominal $\pi^{0}$
mass~\cite{Zyla:2020zbs}. When multiple $\pi^{0}$ candidates are
reconstructed, we retain the one with the smallest $\chi^{2}$ of the
mass-constraint kinematic fit.
 
Furthermore, we require the $p e^{-}e^{-}$ invariant mass to be within
$(1.169, 1.209)$~${\rm{GeV}}/c^{2}$ and the $p \pi^{0}$ invariant mass
to be within $(1.167, 1.201)$~${\rm{GeV}}/c^{2}$.
For the $\Sigma^{-} \to \Sigma^{+}X$ channel, to further suppress
background, one additional kinematic variable is defined as 
\begin{equation}
\vec{\mathbf{p}}_{\rm{miss}} =  \vec{\mathbf{p}}_{\rm{J/\psi}} - \vec{\mathbf{p}}_{\bar \Lambda} - \vec{\mathbf{p}}_{\pi^{+}} - \vec{\mathbf{p}}_{\Sigma^{+}}
\end{equation}
where $\vec{\mathbf{p}}$ are the corresponding momenta in the
$J/\psi$'s center-of-mass system, and $|\vec{\mathbf{p}}_{\rm{miss}}|$
is required to be less than 0.1~${\rm{GeV}}/c$.

Since the $X$ particles are not detected, the DT efficiency of
$\Sigma^{-} \to \Sigma^{+} X$ is only affected by the reconstruction
of the $p$ and $\pi^{0}$ from the $\Sigma^{+}$ decays. Simulation studies
show that due to the limited phase space and small $\Sigma^{-}$
momentum, the momenta and the angular distributions of $p$ and
$\pi^{0}$ are almost the same when $X$ represents different final
states. Therefore, we use the MC samples of $\Sigma^{-} \to \Sigma^{+}
e^{-} e^{-}$ to estimate the efficiency of $\Sigma^{-} \to \Sigma^{+}
X$.

\begin{figure}[!htbp]
\centering
\includegraphics[width=0.38\textwidth]{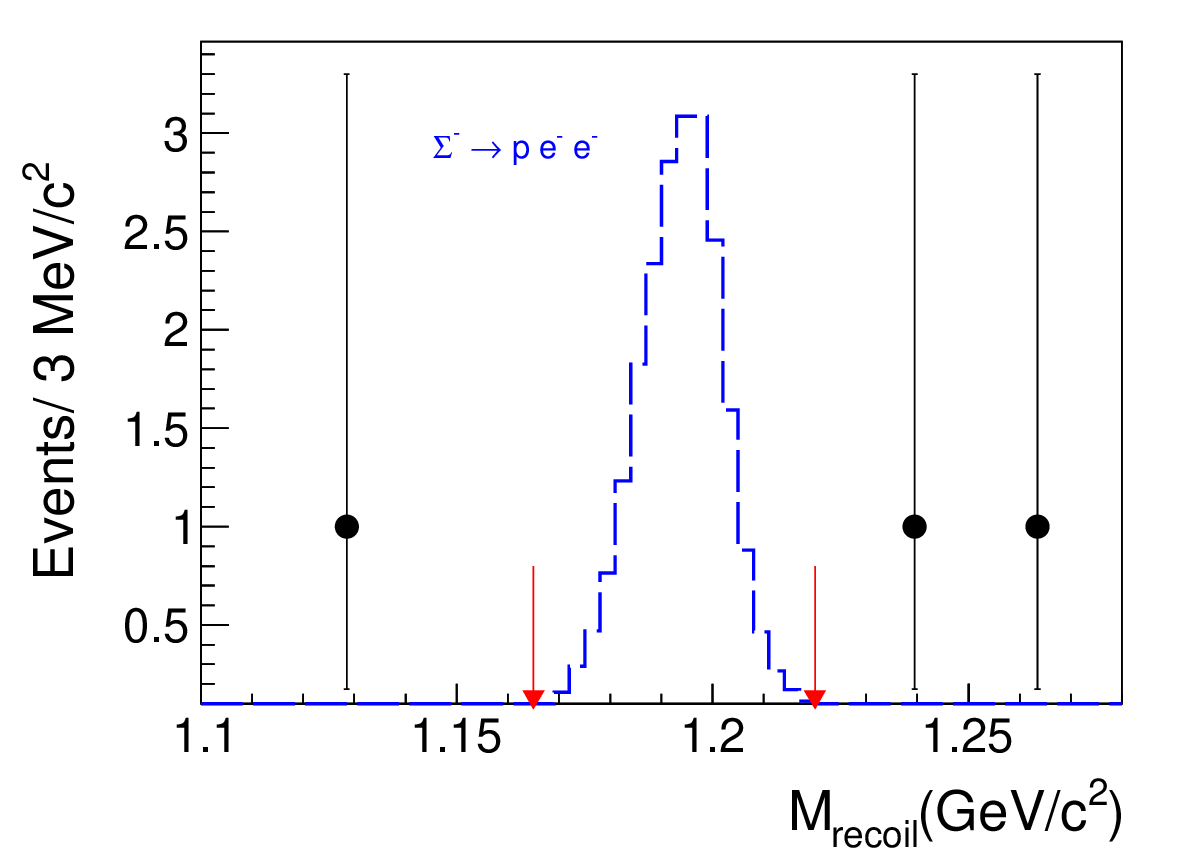}
\includegraphics[width=0.38\textwidth]{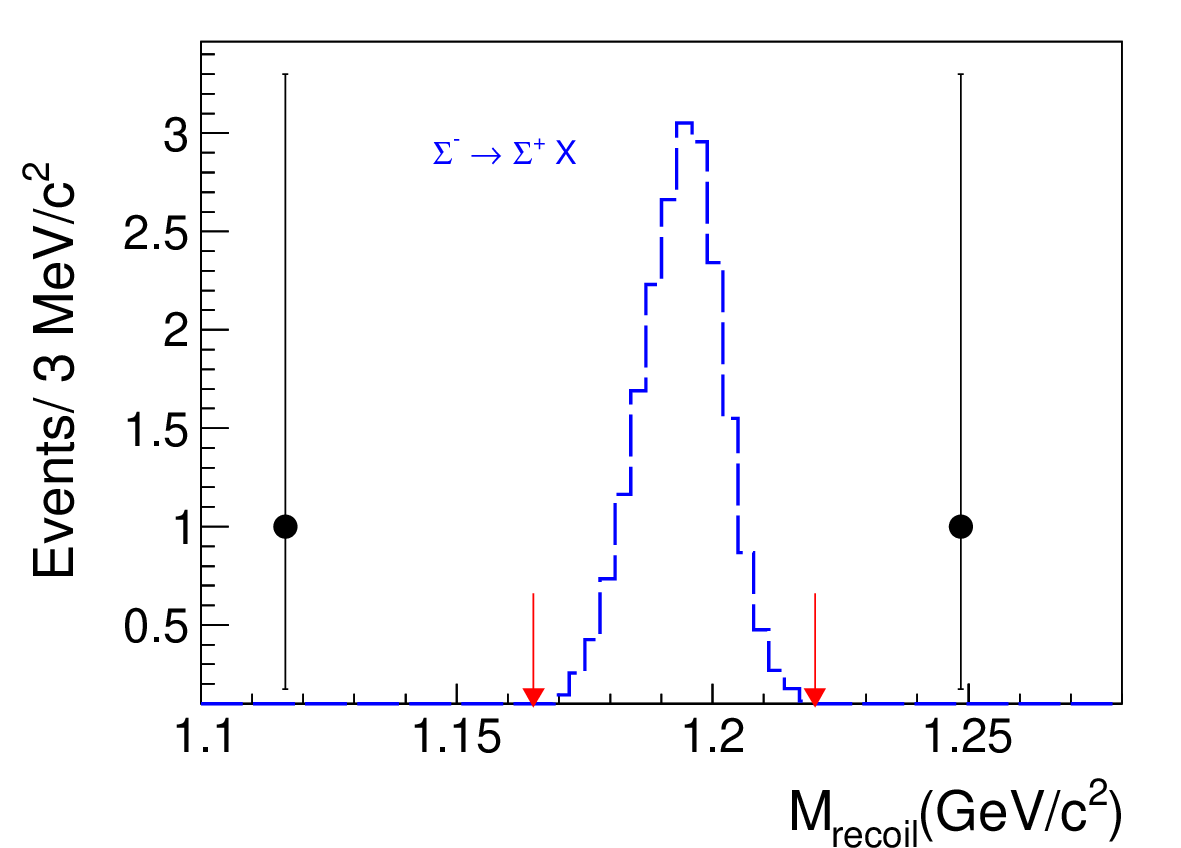}
\caption{$M_{\rm{recoil}}$ distributions of (top) $\Sigma^- \to p e^-
  e^-$ and (bottom) $\Sigma^- \to \Sigma^+ X$. Points with
  error bars are data, and dashed histograms are signal MC simulations
  with arbitrary normalization. The arrows show the signal region.}
\label{fig:result}
\end{figure}

To determine the DT yield, we search for candidates in the
$M_{\rm{recoil}}$ distributions for $\Sigma^- \to p e^- e^-$ and
$\Sigma^- \to \Sigma^+ X$ in data, shown in Fig.~\ref{fig:result}. The
signal region is defined as $[1.165, 1.220]$~${\rm{GeV}}/c^{2}$, which
covers more than $99.7\%$ of all signal events. The DT efficiencies
obtained from MC simulation are 9.02\% and 11.08\%, respectively. No event is
observed in the signal region for either channel.

\subsection{\boldmath Background Study}
Potential background candidates come from the continuum process and
from other $J/\psi$ decays.  To estimate the first kind, we study the
continuum process with data samples collected at $\sqrt{s} = 3.08$, $3.65$, $3.773
~\rm{GeV}$, where the integrated luminosity values are about
150~$\rm{pb}^{-1}$, 50~$\rm{pb}^{-1}$ and 2.93~$\rm{fb}^{-1}$,
respectively. There is no ST peaking background, and no
event passes the DT selection.

\begin{figure}[!htbp]
\centering
\includegraphics[width=0.38\textwidth]{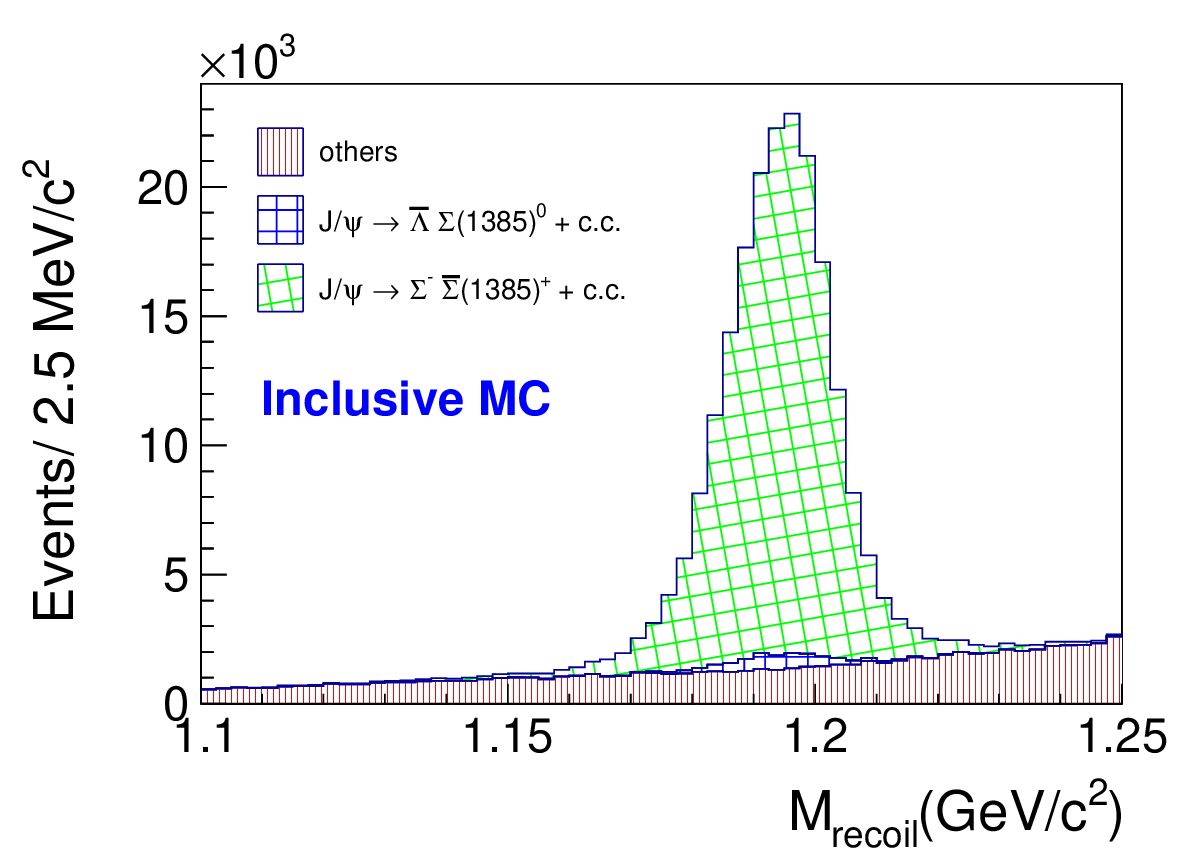}
\caption{$M_{\rm{recoil}}$ distribution of the inclusive Monte Carlo sample.}
\label{fig:recoilingtopo}
\end{figure}

We use the inclusive MC sample to estimate backgrounds from $J/\psi$
decays. The ST peaking background component is from $J/\psi \to
\Sigma(1385)^0 \bar{\Lambda} + c.c.$, and the number of events
scaled to data accounts for about 0.07\% of the total ST yield, so it
is ignored. The smooth background components can be described by a
second order Chebychev polynomial. Figure \ref{fig:recoilingtopo}
shows the different components from the inclusive MC sample. For the
DT selection, only 4 and 3 background events survive in the signal
regions of the $\Sigma^- \to p e^{-} e^{-}$ and $\Sigma^- \to \Sigma^+
X$ channels, respectively, corresponding to normalized numbers of 0.3 and 0.6
background events.

\section{\boldmath SYSTEMATIC UNCERTAINTIES}
The systematic uncertainties in the measurements, summarized in
Table~\ref{tab:uncertainty}, mainly originate from differences between
data and simulation in the tracking and PID efficiency, the tag bias,
the MC model and the cited branching fractions.

The systematic uncertainty due to the proton tracking efficiency is
determined to be $1.0\%$ for each track by studying the two control
samples of $J/\psi \to p K^{-} \bar{\Lambda} + c.c.$ and $J/\psi \to
\Lambda \bar{\Lambda}$~\cite{Ablikim:2012bw}. The uncertainty arising
from the proton PID efficiency is determined with the control sample
$J/\psi \to p \bar{p} \pi^{+} \pi^{-}$. We bin events in the sample in
$\cos\theta$ $(i)$ and $|\vec{\mathbf{p}}|$ $(j)$ of the
proton~\cite{Ablikim:2018uiq} and add differences between data and MC samples
together with the following formula
\begin{equation}
   \Delta \epsilon^{PID} = \sum_{i,j} (\Delta \epsilon^{PID}_{ij} \times \omega^{PID}_{ij}),
\end{equation}
where $\Delta \epsilon^{PID}_{ij}$ is the difference of PID efficiency
and $\omega^{PID}_{ij}$ is the weight factor. The weight factor is
defined as the ratio of the number of events in the $ij$ bin to the
total number of events of the sample.  We use the total differences as
the uncertainties in $\Sigma^{-} \to p e^{-} e^{-}$ and $\Sigma^{-}
\to \Sigma^{+} X$, which are $0.4\%$ and $0.3\%$, respectively.

The uncertainties due to the tracking and PID efficiencies of the
electron are studied with the control sample $e^{+} e^{-} \to \gamma
e^{+} e^{-}$ (at $\sqrt{s} = 3.097$~$\rm{GeV}$). Similar as above, we bin
events in the sample by $\cos\theta$ and $|\vec{\mathbf{p}}_{t}|$
$\left ( |\vec{\mathbf{p}}| \right )$ for tracking (PID). The
uncertainties of the tracking efficiency for the high-momentum and
low-momentum electrons are $0.5\%$ and $3.1\%$, respectively, and the
uncertainties of the PID efficiency are $0.6\%$ and $2.2\%$,
respectively. The total differences for tracking and PID for
$\Sigma^{-} \to p e^{-} e^{-}$ are $3.6\%$ and $2.8\%$,
respectively. The uncertainties associated with photon detection and
$\pi^{0}$ reconstruction are obtained from the control sample $J/\psi
\to \pi^{+} \pi^{-} \pi^{0}$~\cite{Ablikim:2010zn}. The differences
between data and MC samples are $1.0\%$ per photon and $1.0\%$ per $\pi^{0}$,
respectively.

The systematic uncertainty due to the $\Sigma^{+}(p \pi^{0})$ mass
window is determined to be $0.6\%$ using a control sample of
$J/\psi \to \Sigma^{+} \bar{\Sigma}^{-}$
decays~\cite{Ablikim:2012jg}. Since we do not have a sample of
$\Sigma^{-} \to p e^{-} e^{-}$, we change the limits to those
obtained from the mass distribution of $\Sigma^{-}$ whose decay is
determined by a phase space model. The relative change of the DT efficiency,
$0.5\%$, is taken as the uncertainty of $\Sigma^{-}(pe^{-}e^{-})$ mass
window.

The main uncertainties in the ST selection, including the total number of
$J/\psi$ events, the reconstruction of $\bar{\Lambda}$ and the
bachelor $\pi^{+}$, cancel in the double tag method. The tag
bias is related with the MC sample used to obtain the ST efficiency. We
change the sample with the decay chain $J/\psi \to {\bar
  \Sigma(1385)^+} \Sigma^-(\Sigma^- \to X)$ to the sample with the
decay chain $J/\psi \to {\bar \Sigma(1385)^+} \Sigma^-(\Sigma^- \to p
e^{-}e^{-})$ and $J/\psi \to {\bar \Sigma(1385)^+} \Sigma^-(\Sigma^-
\to \Sigma^{+} e^{-}e^{-})$. The average relative change of the ST
efficiency is taken as the associated uncertainty, which is
$1.3\%$. The statistical uncertainty of the tag yields is $0.38\%$,
and it is taken into account together with the statistical
uncertainties of efficiencies when calculating the ULs.

To estimate the uncertainty of the MC model for the signal, we change the
values of parameters in the model that describes the $q^{2}$-dependent
differential decay width of $B_{1}^{-} \to
B_{2}^{+}l^{-}l^{-}$~\cite{Barbero:2013fc}. We take the relative
changes of the DT
efficiencies as the associated uncertainties for $\Sigma^{-} \to p
e^{-} e^{-}$ and $\Sigma^{-} \to \Sigma^{+} X$, which are $1.0\%$ and
$0.9\%$, respectively.

Since the limit for $|\vec{\mathbf{p}}_{\rm{miss}}|$
(0.1~${\rm{GeV}}/c$) is much larger than the kinematic limit, the
uncertainty of the requirement on $|\vec{\mathbf{p}}_{\rm{miss}}|$ is negligible. The
relative uncertainties for branching fractions of $\pi^{0} \to
\gamma \gamma$ and $\Sigma^{+} \to p \pi^{0}$ are taken from the
PDG~\cite{Zyla:2020zbs}, and are 0.034\% and 0.6\%,
respectively. The uncertainty for the $\pi^{0}$ is so small that it
can be ignored. The total systematic uncertainties are obtained by
adding all uncertainties above in quadrature.

\begin{table}[!htbp]
  \centering
  \caption{The relative systematic uncertainties (in \%) on the
    branching fraction measurements.}
  \begin{tabular}{p{3.8cm}p{2.2cm}<{\centering}p{2.2cm}<{\centering}}
  \hline 
  \hline
   Source                   &  $\Sigma^- \to p e^- e^-$  &  $\Sigma^- \to \Sigma^+ X$  \\    
  \hline
   Tracking of proton      &   1.0    &   1.0     \\
   PID of proton           &   0.4    &   0.3     \\    
   Tracking of electron    &   3.6    &    $\cdots$      \\
   PID of electron         &   2.8    &    $\cdots$      \\
   Photon detection   &      $\cdots$       &    2.0   \\
   $\pi^0$ reconstruction   &    $\cdots$    &   1.0     \\
   $\Sigma^{-}$ mass window   &  0.5   & $\cdots$    \\
   $\Sigma^{+}$ mass window  &  $\cdots$ &  0.6  \\
   Tag bias                 &   1.3    &   1.3     \\ 
   MC model          &   1.0   &   0.9     \\
   Quoted branching ratios   &    $\cdots$   &   0.6    \\
  \hline
   Total                    &   5.0   &   3.1      \\   
  \hline
  \hline
  \end{tabular}     
  \label{tab:uncertainty}  
\end{table}

\section{\boldmath RESULTS}
The ULs for the signal yields are calculated using a frequentist method
with an unbounded profile likelihood treatment of systematic
uncertainties, which is implemented by the class {\sc TROLKE} in the
ROOT framework~\cite{Rolke:2004mj}. The number of the signal and
background events are assumed to follow a Poisson distribution, the
detection efficiency is assumed to follow a Gaussian distribution, and
the systematic uncertainty is considered as the standard deviation of
the efficiency. The resulting UL for the branching fraction is determined
by
\begin{equation}
\mathcal{B}^{up}_{\rm{sig}} < \frac{s_{90}^{\rm{obs}}}{N_{\rm{tag}}^{\rm{obs}}\epsilon_{\rm{DT}} /\epsilon_{\rm{ST}}},
\label{equ:UP}
\end{equation}
where $s_{90}^{\rm{obs}}$ is the upper limit on the number of signal events determined at the
$90\%$ CL. The ULs for branching fractions are
\begin{gather}
\nonumber
   \mathcal{B} \left( \Sigma^{-} \to p e^{-} e^{-} \right )  <  6.7 \times 10^{-5}, \\
\nonumber
   \mathcal{B} \left( \Sigma^{-} \to \Sigma^{+} X  \right )  < 1.2 \times 10^{-4}.
\end{gather}

\section{\boldmath SUMMARY}
To summarize, with the data sample of $(1310.6 \pm 7.0) \times 10^{6}$
$J/\psi$ events collected by BESIII detector, a search for the LNV
decay $\Sigma^{-} \to p e^{-} e^{-}$ and the rare inclusive decay
$\Sigma^{-} \to \Sigma^{+} X$ is performed for the first time. No
signal event is observed, and the upper limits on the branching
fractions of $\Sigma^{-} \to p e^{-} e^{-}$ and $\Sigma^{-} \to
\Sigma^{+} X$ at the $90\%$ CL are $6.7 \times 10^{\ -5}$ and $ 1.2
\times 10^{-\ 4}$, respectively. Our results are well above the
prediction in references~\cite{Barbero:2002wm, Barbero:2007zm,
  Barbero:2013fc}. 

\section*{\boldmath ACKNOWLEDGEMENTS}
The BESIII collaboration thanks the staff of BEPCII and the IHEP computing center for their strong support. This work is supported in part by National Natural Science Foundation of China (NSFC) under Contracts Nos. 11625523, 11635010, 11735014, 11822506, 11835012, 11935015, 11935016, 11935018, 11961141012, 12035009; the Chinese Academy of Sciences (CAS) Large-Scale Scientific Facility Program; Joint Large-Scale Scientific Facility Funds of the NSFC and CAS under Contracts Nos. U1532257, U1732263, U1832207; CAS Key Research Program of Frontier Sciences under Contracts Nos. QYZDJ-SSW-SLH003, QYZDJ-SSW-SLH040; 100 Talents Program of CAS; INPAC and Shanghai Key Laboratory for Particle Physics and Cosmology; ERC under Contract No. 758462; German Research Foundation DFG under Contracts Nos. 443159800, Collaborative Research Center CRC 1044, FOR 2359, FOR 2359, GRK 214; Istituto Nazionale di Fisica Nucleare, Italy; Ministry of Development of Turkey under Contract No. DPT2006K-120470; National Science and Technology fund; Olle Engkvist Foundation under Contract No. 200-0605; STFC (United Kingdom); The Knut and Alice Wallenberg Foundation (Sweden) under Contract No. 2016.0157; The Royal Society, UK under Contracts Nos. DH140054, DH160214; The Swedish Research Council; U. S. Department of Energy under Contracts Nos. DE-FG02-05ER41374, DE-SC-0012069.


\end{document}